%% file: main.tex
\documentclass[11pt]{article}
\usepackage[preprint]{acl}
\usepackage{times,latexsym}
\usepackage[T1]{fontenc}
\usepackage[utf8]{inputenc}
\usepackage{microtype,inconsolata}
\usepackage{amsmath,amssymb,booktabs,array,graphicx}
\usepackage{float,placeins}
\usepackage{sivia-style}
\floatstyle{ruled}
\newfloat{algorithm}{t}{loa}
\floatname{algorithm}{Algorithm}
\newcommand{\sivia}{Sivia}
\newcommand{\base}{Brief}
\newcommand{\plan}{Plan}
\newcommand{\strong}{Reference}
\newcommand{\RRA}{\operatorname{RRA}}
\newcommand{\repoURL}{https://github.com/exsinger-hub/Sivia}
\title{Sivia: Learning to Draw, Learning to Learn}
\author{%
{\normalsize
Feng Yuan\textsuperscript{1},
Xinyi Gao\textsuperscript{2},
Zijun Sun\textsuperscript{3},
Haoyue Li\textsuperscript{1},
Menglei Hua\textsuperscript{4},
Xiao Li\textsuperscript{5},
Yifan Gao\textsuperscript{1}}%
\\[5pt]
{\small\normalfont
\textsuperscript{1}USTC
\quad
\textsuperscript{2}University of Manchester
\quad
\textsuperscript{3}University of Zurich}%
\\[2pt]
{\small\normalfont
\textsuperscript{4}Harbin Medical University
\quad
\textsuperscript{5}Beihang University}%
}
\hypersetup{pdfauthor={Feng Yuan, Xinyi Gao, Zijun Sun, Haoyue Li, Yifan Gao, Menglei Hua, Xiao Li}}
\begin{document}
\raggedbottom
\maketitle
\begin{abstract}
A paper's overview figure is often the first thing readers see: it must explain the method correctly and look polished. We present \sivia{}, an overview-figure system that reads the paper, plans its relations, adapts visual organization from recent top-conference papers, and composes a design for image generation. \sivia{} improves with use through human-guided recursive self-improvement (RSI) at two levels: \emph{learning to draw} revises its drawing skill from feedback on accuracy and presentation, and \emph{learning to learn} revises the updater that proposes those revisions. A personal profile adapts presentation to its user's aesthetic preferences while the scientific content stays fixed. A diagnostic study of 1,964 overviews for 445 papers (17,973 relation judgments) shows that most errors are omissions and that source-derived relation plans raise required-relation accuracy (RRA) by up to 19.85 points. On 48 test papers, learning both levels reaches 93.54\% RRA, 15.31, 9.38, and 8.33 points above the initial skill, the Fixed-updater, and the GEPA baselines, while readability and visual quality rise from 3.95 and 4.00 to 4.70 and 4.78 on five-point scales. Learned presentation profiles raise liking from 4.02 to 4.73 and adoption from 34 to 45 of 48 tasks. Code and released artifacts are available at \url{\repoURL}.
\end{abstract}
\suppressfloats[t]
\begin{figure}[t]
\centering
\includegraphics[width=\columnwidth, trim=0 94.5 51.5 0, clip]{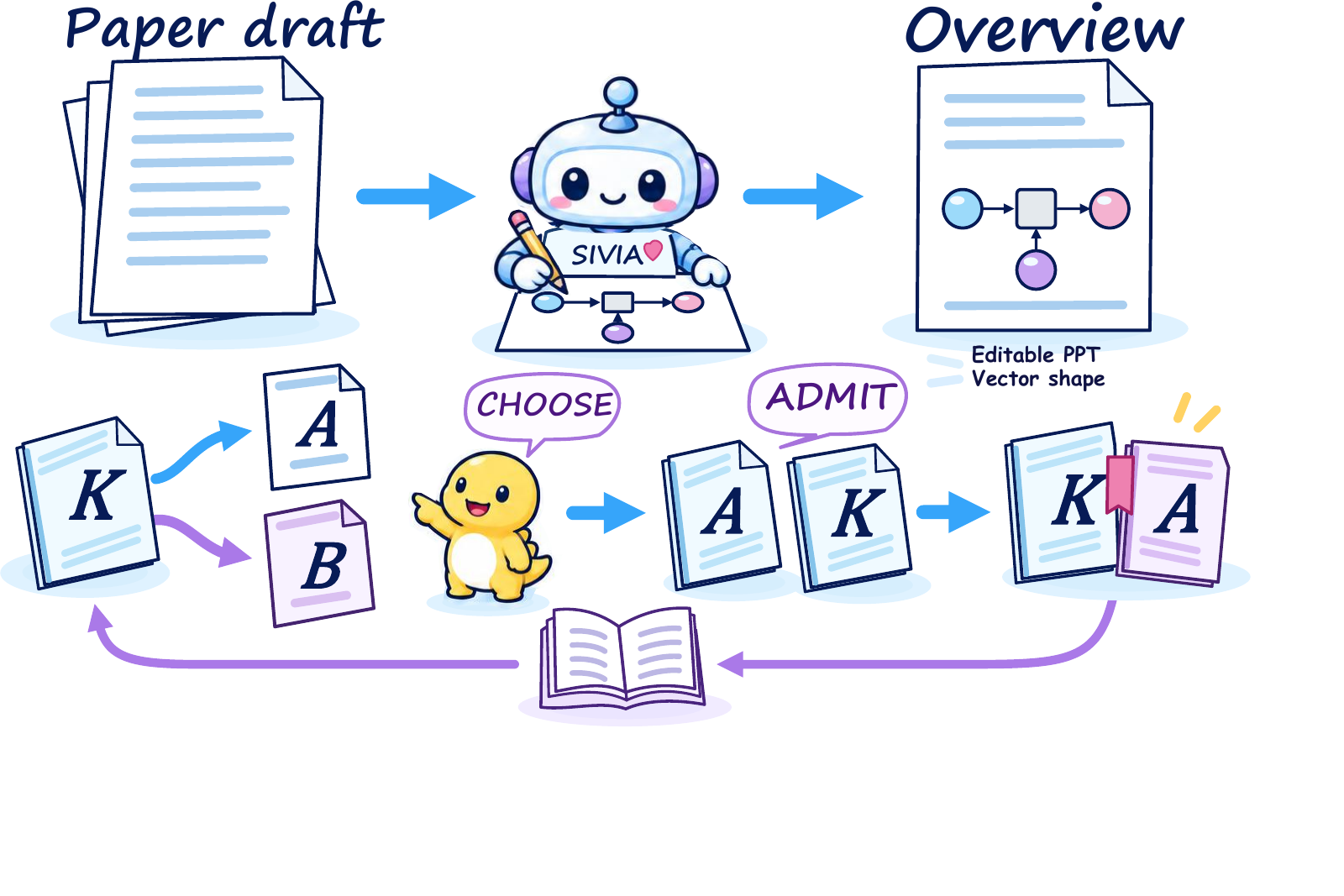}
\caption{Sivia turns papers into overview figures. The lower loop schematizes human selection and admission; Figure~\ref{fig:sivia-method} details the coupled skill/updater update.}
\label{fig:sivia-graphical-abstract}
\end{figure}
\input{sections/introduction}
\input{sections/related}
\input{sections/diagnostic}
\input{sections/method}
\input{sections/experiments}
\input{sections/conclusion}
\bibliography{references}
\clearpage
\appendix
\setcounter{dbltopnumber}{4}
\renewcommand{\dbltopfraction}{0.9}
\input{sections/appendix}

\end{document}

%% file: sections/introduction.tex
\section{Introduction}
\label{sec:intro}
The overview figure is the visual core of a paper. It is often the first thing reviewers and readers examine, and it shapes how they understand the method: which component consumes which artifact, in which stage, and under which condition. A good overview must therefore be both accurate and beautiful, with the clarity and polish expected at top venues. Image generators now render attractive diagrams from text \citep{zala2024diagrammergpt,zhu2026paperbanana}, yet an attractive figure can still route a verifier to the wrong input or drop a perception step, and an accurate figure can still be too dense or plain to communicate.

To locate such failures, we assess 1,964 first-draft overviews of 445 papers against 4,033 source-anchored relation requirements, totaling 17,973 human judgments by three raters (Section~\ref{sec:diagnostic}). Omitted relations account for 63\% of non-correct judgments, and a relation plan extracted from the target paper raises required-relation accuracy (RRA) by up to 19.85 points, whereas a retrieved reference figure does not. Accuracy and appearance can also pull apart: in one cohort, plans make figures more complete but less readable.

Improving such figures over time is not straightforward. A lesson from one paper, such as \emph{draw the observation that follows each state-changing action}, must transfer as a question to ask rather than as content to copy, because similar-looking operations in another paper may consume different inputs. Aesthetic judgments are contextual and personal: what looks right depends on the figure's purpose and on its author's taste. The procedure that turns feedback into instruction edits is itself a design choice; prompt optimizers fix it \citep{agrawal2025gepa,pryzant-etal-2023-automatic}, and Promptbreeder and the Darwin G\"odel Machine evolve it \citep{fernando2024promptbreeder,zhang2025darwin}, but without a criterion of fidelity to each new source.

We present \sivia{}, an overview-figure system that becomes more accurate and more attractive with use (Figure~\ref{fig:sivia-graphical-abstract}). It reads the paper, plans source-grounded relations, adapts the visual organization of overview figures from recent top-conference papers, and composes a detailed design for the image generator. Its state consists of a drawing skill $K$, an updater $U$, and a history $H$ of edits and their outcomes. \emph{Learning to draw}: the updater converts human findings on accuracy and presentation into bounded skill edits. \emph{Learning to learn}: a meta-proposal revises the updater from the outcomes of earlier edits; the current and proposed updaters edit the same parent skill, and humans select and admit a skill together with its producing updater. A personal profile $P_u$ further adapts presentation to its user's aesthetic preferences while the scientific content stays fixed.

On 48 test papers, learning both levels reaches 93.54\% RRA, 15.31, 9.38, and 8.33 points above the initial skill, the Fixed-updater, and the GEPA baselines, and raises readability and visual quality from 3.95 and 4.00 to 4.70 and 4.78 on five-point scales. From a shared parent skill, the learned updater's proposal exceeds that of the initial updater by 9.90 points, and learned presentation profiles raise liking from 4.02 to 4.73. Our contributions are:
\begin{itemize}
  \setlength{\itemsep}{1pt}
  \setlength{\parskip}{0pt}
  \item \sivia{}, an overview-figure system that targets accuracy and aesthetics together, combining source-grounded relation planning with detailed conference-style visual design;
  \item human-guided recursive self-improvement (RSI) at two levels, in which \sivia{} improves both its drawing skill and the procedure that improves it: skill edits phrased as transferable reading rules, an updater revised from the outcomes of earlier edits under same-parent comparison and two-stage human admission, and a personal profile that adapts presentation to its user;
  \item a 445-paper diagnostic study and an evaluation that separately measures relation accuracy, error types, readability, visual quality, and preference.
\end{itemize}

%% file: sections/related.tex
\section{Related Work}
\paragraph{Overview and scientific figure generation.}
DiagrammerGPT plans layouts before synthesis \citep{zala2024diagrammergpt}, and SciDoc2Diagrammer-MAF combines document understanding with multi-aspect feedback \citep{mondal2024scidoc}. PaperBanana, whose implementation is now named PaperVizAgent (PVA), integrates retrieval, planning, styling, visualization, and criticism \citep{zhu2026paperbanana}. FigAgent evolves reusable drawing middleware \citep{li2026figagent}, and SciDiagramEdit learns editing skills from paper revisions \citep{sun2026scidiagramedit}. These systems improve individual figures or components; \sivia{} learns instructions for recovering a later paper's roles and dependencies, and learns how to revise them.
\begin{figure*}[t]
\centering
\includegraphics[width=0.94\textwidth]{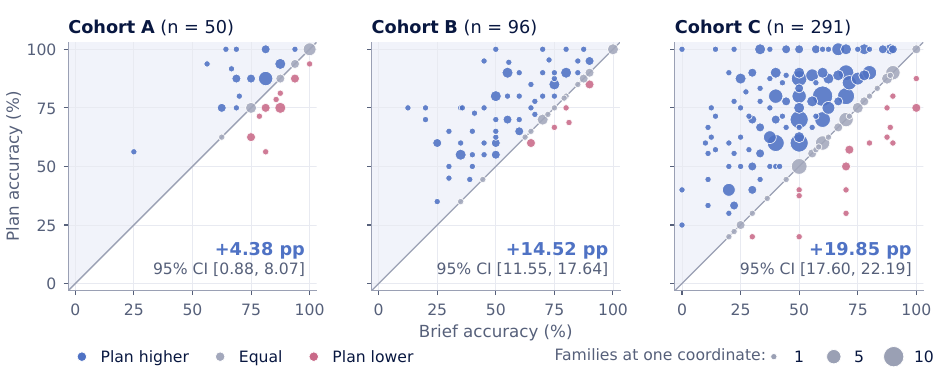}
\caption{Diagnostic study: family-level RRA with a source-derived relation plan (\plan{}) versus the brief alone (\base{}). Points above the diagonal favor \plan{} (A: 24 higher / 12 equal / 14 lower; B: 70/19/7; C: 223/49/19); marker area counts coincident families. Labels give paired mean gains with 95\% family-bootstrap intervals; $n$ counts families with all prescribed outputs.}
\label{fig:scatter}
\end{figure*}

\paragraph{Learning from feedback and learning to improve.}
Self-Refine iterates on the current output \citep{madaan2023selfrefine}, and Reflexion stores verbal feedback for later episodes \citep{shinn2023reflexion}. ProTeGi, DSPy, TextGrad, and GEPA optimize prompts or pipeline components from critiques, metrics, or execution traces \citep{pryzant-etal-2023-automatic,khattab2024dspy,yuksekgonul2025textgrad,agrawal2025gepa}; ACE curates strategies in an evolving playbook \citep{zhang2025agenticcontextengineeringevolving}, and SEAL learns to generate self-edits \citep{zweiger2025seal}. Promptbreeder evolves mutation prompts alongside task prompts, the Darwin G\"odel Machine evolves self-modifying agents, and Hyperagents makes the modifying meta agent editable \citep{fernando2024promptbreeder,zhang2025darwin,zhang2026hyperagents}; learning the improvement procedure, the core of recursive self-improvement, is therefore not new in itself. These methods select by task fitness on a fixed task distribution. In scientific illustration, each new paper defines a different correct structure, so an edit that helps one paper can mislead on another. \sivia{} addresses this setting with three specific mechanisms: skill edits are phrased as reading rules whose answers come from the new source; competing updaters edit the same parent from the same findings, so that proposal quality can be compared directly; and the updater is revised from an edit-level history of earlier changes, their outcomes, and human admission decisions, including rejected edits.

\paragraph{Relation-level evaluation and personalized aesthetics.}
Scientific information extraction recovers entities and relations from papers \citep{luan2018scierc,jain-etal-2020-scirex}. Fine-grained text-to-image metrics decompose prompts into verifiable requirements \citep{hu2023tifa,cho2024dsg,ghosh2023geneval}, and SciFlow-Bench evaluates diagram structure through inverse parsing \citep{zhang2026sciflow}. We likewise score source-anchored relation requirements, distinguishing omission, ambiguity, and incorrectness, and keep presentation ratings separate. Our preference extension relates to profile-based personalization of language models \citep{salemi-etal-2024-lamp,alhafni-etal-2024-personalized}, applied here to presentation under a fixed scientific contract.

%% file: sections/diagnostic.tex
\section{What Goes Wrong in Generated Overviews}
\label{sec:diagnostic}
\label{sec:task}
\begin{figure*}[t]
\centering
\includegraphics[width=\textwidth]{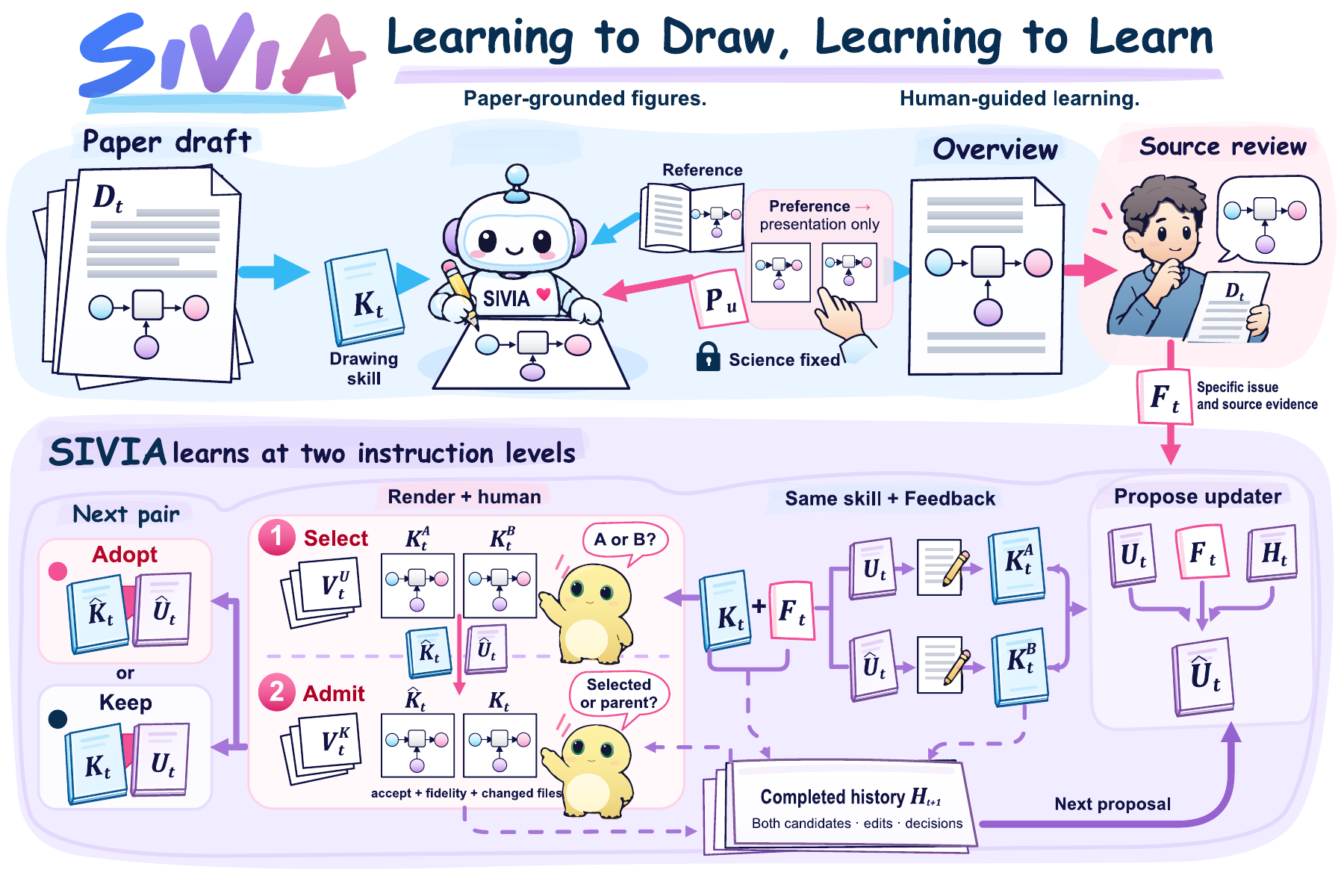}
\caption{Overview of \sivia{}. Top: source-grounded drawing produces figures for human review. Bottom: the current and a proposed updater edit the same parent skill from shared findings; human comparisons select a candidate and decide whether to admit its skill together with its producing updater. Completed edit--policy--outcome records inform later proposals (Algorithm~\ref{alg:rsi}). Personal preferences affect presentation under a fixed scientific contract.}
\label{fig:sivia-method}
\end{figure*}
\paragraph{Task and measure.}
Given a paper $x$ and an overview scope $s$, a drawing workflow produces an image $I=G(C(x,s))$, where $C$ plans the figure and writes the production prompt and $G$ is an image generator. A scientific relation specifies participating roles, the artifact or operation connecting them, its direction, and any source-supported condition or phase; building code from a plan and checking code against that plan involve the same entities but different relations. For a frozen checklist $Q_f$ of family $f$ and its $R_f$ evaluated images, required-relation accuracy is
\begin{equation}
\RRA_f=\frac{1}{R_f}\sum_{r=1}^{R_f}
\frac{\sum_{q\in Q_f}\mathbf{1}[y_{frq}=\text{correct}]}{|Q_f|}.
\label{eq:rra}
\end{equation}
Non-correct labels are omitted, ambiguous, or incorrect; the learning evaluation additionally separates severe errors (Section~\ref{sec:setup}). Scores average repetitions within a family and then weight families equally; a family groups all images generated for one paper.

\paragraph{Diagnostic corpus.}
We select 445 papers from over 5,000 candidates, focusing on recent leading-venue research on agents, retrieval, reasoning, verification, and learning methods. Three cohorts compare generation inputs within each family: \base{} supplies an overview brief; \strong{} adds a retrieved external figure; \plan{} adds a relation plan extracted from the target paper without access to any ratings; Cohort~B also includes a random reference. Of 1,985 scheduled native image requests, 1,964 return images, with no retry or selection. Before generation, each family receives 5--14 source-anchored requirements; three raters label every image against them, blind to condition, yielding 17,973 judgments over 4,033 requirements (Appendix~\ref{app:diagnostic}). Cohorts A, B, and C contain 50, 100, and 295 families, with two, two, and one outputs per condition.

\paragraph{Annotation protocol.}
Requirements are drafted before generation by a separate language-model session, anchored to source passages, and specify the minimum visible evidence; compound items count as correct only when all components are visible. Relation-plan extraction is separate from checklist preparation and sees neither checklists nor ratings. Retrieval for \strong{} uses TF--IDF in Cohort~A and \texttt{all-mpnet-base-v2} embeddings in B and C over pools of 948 and 781 papers, excluding the target and its aliases. During review, conditions and prompts are hidden, and paper blocks and images are shuffled. Raters may object to a requirement's source support; objections affect 30 requirements in 27 families, and excluding them changes the \plan{}$-$\base{} gains only to +4.26, +14.38, and +19.85 points.

\paragraph{Relation plans help; adding a reference does not.}
Explicit relation planning improves RRA in all three cohorts, by +4.38, +14.52, and +19.85 points (Figure~\ref{fig:scatter}). Conditioning on a retrieved reference figure instead lowers RRA by 4.03, 2.36, and 2.31 points, and in Cohort~B retrieved and random references differ by only +0.05 points $[-1.78,1.88]$. A reference figure informs visual organization but does not supply the target's relations.

\paragraph{Omission is the dominant failure.}
Omissions account for 4,104 of the 6,530 non-correct judgments (62.85\%). Planning reduces the omission rate from 26.31\% to 16.73\% in Cohort~B and from 27.86\% to 16.05\% in Cohort~C, which accounts for 66.0\% and 59.5\% of the respective gains. Presentation is a separate objective: in Cohort~B the plan condition's readability is 0.19 points lower on a five-point scale.

\paragraph{Failures arise at two stages.}
Tracing requirements from source to plan to prompt to image separates two failure stages, illustrated here by two selected cases (Appendix~\ref{app:diagnostic-cases}). For Mobile-Agent-RAG \citep{zhou2026mobileagentrag}, the plan and prompt both require perception after each mobile action, yet one of two images from the same prompt omits it: a \emph{realization} failure. For AI-Researcher \citep{tang2025airesearcher}, checking code against the implementation plan never enters the extracted plan: a \emph{planning} failure. The two cases call for different instruction changes, namely stating endpoints and phases explicitly, or extracting relations more completely.

%% file: sections/method.tex
\section{Sivia}
\label{sec:method}
\sivia{} couples a source-grounded drawing workflow with two levels of instruction learning (Figure~\ref{fig:sivia-method}). A drawing skill $K$ is a set of text files that governs scientific abstraction, relation planning, and visual communication; an updater $U$ is a text policy that instructs a fixed language model to turn feedback into skill edits; a history $H$ records completed edits, their producing updaters, and their outcomes. Learning changes only $K$ and $U$; model weights remain fixed.

\subsection{Source-grounded drawing}
\label{sec:drawing}
Guided by $K$, the assistant reads the target paper and records a source-linked contract: nodes, roles, operator identities, phases, directed relations, and required input sets, each anchored to source passages. Shared operator identity distinguishes reuse of one operation across stages from unrelated boxes with similar labels. The assistant then retrieves candidate figures from a reference library built from overview figures of recent top-conference papers, each stored with its complete design prompt, and inspects these prompts and images; a reference contributes visual organization only, consistent with Section~\ref{sec:diagnostic}. Finally, it composes a detailed production prompt that specifies hierarchy, layout, palette, typography, and iconography in a conference style, from the contract, the selected template, and any applicable user preferences, and the native image generator renders a draft for review at publication size. Instruction traces link each applied rule and requirement to a prompt span. We write the drawing workflow for paper $x$ as $\mathcal{R}(x,K)$. Each candidate derives its contract from the permitted source with its own skill; evaluation checklists are never visible to the drawing actor, so learned skills cannot encode the test requirements.

\subsection{Learning to draw: skill updates}

\paragraph{Grounded findings.}
Each iteration $t$ uses three prescheduled, disjoint batches of paper families: discovery $D_t$, updater selection $V_t^U$, and skill admission $V_t^K$, with 16, 8, and 8 families in our experiments. Human findings $F_t$ describe visible issues in the discovery figures, including scientific fidelity, scope, mechanism coverage, reading path, and legibility. Scientific findings cite passages and pages of the source paper, so that a reported consumer, direction, or condition error is grounded in evidence rather than in taste.

\begin{algorithm}[t]
\caption{Skill and updater learning.}
\label{alg:rsi}
\begin{minipage}{0.97\columnwidth}
\small
\textbf{Input:} $K_0,U_0,H_0$; disjoint batches
$\{D_t,V_t^U,V_t^K\}_{t=0}^{T-1}$; fixed models.
\begin{enumerate}
  \setlength{\itemsep}{2pt}
  \setlength{\parskip}{0pt}
  \item For $t=0,\ldots,T-1$:
  \item \hspace*{0.5em}$F_t \leftarrow \operatorname{HumanFindings}(\mathcal{R}(D_t,K_t))$.
  \item \hspace*{0.5em}$\widetilde{U}_t \leftarrow \mathcal{P}(U_t,F_t,H_t)$.
  \item \hspace*{0.5em}$K_t^A \leftarrow \mathcal{E}(K_t;U_t,F_t)$;
  $K_t^B \leftarrow \mathcal{E}(K_t;\widetilde{U}_t,F_t)$.
  \item \hspace*{0.5em}Compare $K_t^A$, $K_t^B$ blindly on $V_t^U$; humans choose $c_t\in\{A,B\}$.
  \item \hspace*{0.5em}$(\widehat{K}_t,\widehat{U}_t)\leftarrow(K_t^A,U_t)$ if $c_t=A$, else $(K_t^B,\widetilde{U}_t)$.
  \item \hspace*{0.5em}Compare $\widehat{K}_t$ with $K_t$ on $V_t^K$; humans decide admission $a_t$.
  \item \hspace*{0.5em}If $a_t$ accepts, no severe error is found, and $\widehat{K}_t\ne K_t$: $(K_{t+1},U_{t+1})\leftarrow(\widehat{K}_t,\widehat{U}_t)$; else keep $(K_t,U_t)$.
  \item \hspace*{0.5em}$H_{t+1}\leftarrow H_t\cup\{$both edits, policies, judgments, decisions$\}$.
  \item Return the final skill and updater.
\end{enumerate}
\end{minipage}
\end{algorithm}

\paragraph{Bounded edits.}
The edit operator $\mathcal{E}(K;U,F)$ asks the text model, instructed by $U$, to diagnose $F$ and propose replacements. Each edit names an existing skill file and an exact span that must occur once; a proposal contains at most six edits and 12,000 changed characters, and an empty proposal keeps the parent. The edits target questions rather than content. A finding that a perception step was omitted after a mobile action, for example, can yield the rule \emph{for each state-changing action, identify the source-specified observation required before the next decision, and assign it an explicit stage and consumer}. It directs each new paper's reading toward a relation type that is often omitted.

\paragraph{Design rules.}
Edits can equally target presentation, because findings also cover reading path and legibility. A design rule might order stages along a single reading direction, limit the palette to a few semantic colors, or replace a dense text box with an icon and a short label. Like reading rules, design rules apply to every later paper and are judged by readability and visual quality, so the skill accumulates both what to show and how to show it.

\subsection{Learning to learn: updater updates}

\begin{figure*}[t]
\centering
\includegraphics[width=\textwidth]{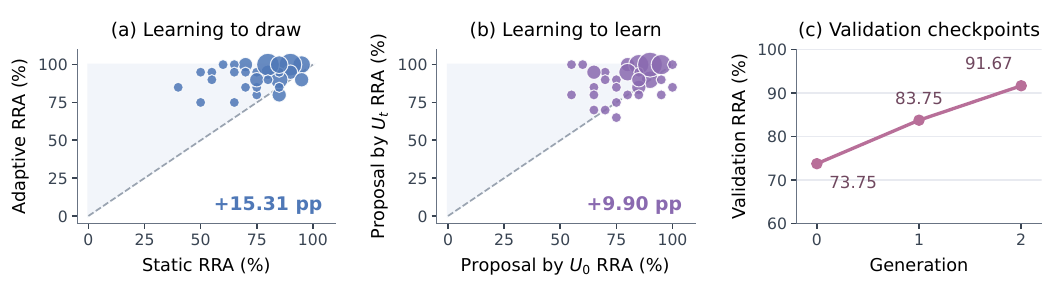}
\caption{Learning effects on the 48 test families. (a) Learning to draw: Adaptive-updater versus the initial skill. (b) Learning to learn: from the same parent skill and findings, the skill proposed by the learned updater $U_t$ versus that proposed by $U_0$. Points above the diagonal favor the learned condition; marker area counts coincident families. (c) Checkpoint RRA at generations 0--2 on the 24 validation families.}
\label{fig:rsi-effects}
\end{figure*}

\paragraph{Outcome-conditioned updater proposal.}
A fixed meta-proposal instruction revises the updater from the current findings and the history, $\widetilde{U}_t=\mathcal{P}(U_t,F_t,H_t)$: the recursion is two-level, revising the skill and the updater but not the instruction that proposes updater edits. The history provides, for every completed iteration, the producing policies, the exact before/after skill spans, human judgments, and admission decisions, including rejected and harmful edits. The proposed updater therefore specifies how to infer transferable rules, where to edit, and how broadly, in light of which earlier edits transferred to later papers.

\paragraph{Same-parent candidates and two-stage admission.}
The current and the proposed updater edit the \emph{same} parent skill from the \emph{same} findings, producing $K_t^A$ and $K_t^B$ (Algorithm~\ref{alg:rsi}). Their difference is thus attributable to the updaters for that parent and those findings. Raters first compare the two candidates on $V_t^U$ through anonymized image pairs and their sources, recording scientific assessments and visual preferences before making an explicit choice. Raters then compare the chosen skill against its parent on $V_t^K$. Admission requires explicit acceptance, no severe scientific error, and a change in the skill files; it promotes the skill and its producing updater together, so that the policy that produced a successful edit governs later proposals. Every completed iteration, including rejected candidates, enters the history consumed by the next proposal.

\subsection{Personal aesthetic profiles}
\label{sec:personal}
A user profile $P_u$ stores presentation rules, currently for composition density and illustration abstraction, with their applicability conditions (e.g., purpose, medium, width) and supporting feedback; explicit instructions override them. Profiles change only presentation: the shared skill and the source contract, including node identities, input sets, directions, and stages, stay fixed, and applied rules appear in the production prompt through traces. The profile is learned from use: when the user is satisfied with an output, that figure and its design prompt join the user's reference library, and the profile's presentation rules are refined from the same feedback. Liking, scientific fidelity, and profile admission are assessed separately.

%% file: sections/experiments.tex
\section{Experimental Setup}
\label{sec:setup}
\input{sections/profile-figure}
\paragraph{Data.}
Each learning condition runs two iterations of Algorithm~\ref{alg:rsi} with 16 discovery, 8 selection, and 8 admission families per iteration; 24 validation families monitor checkpoints, and the final evaluation uses 48 disjoint test families that give no feedback to learning. Each condition produces two outputs per test family, and each output is scored against ten source-anchored relation requirements; the ten conditions yield 960 outputs and 9,600 adjudicated relation labels. All text roles in all conditions, including updaters and baselines, use GPT-6.1 Sol, and all images come from the same native image generator.

\paragraph{Conditions.}
\textbf{Direct} generates from the paper without retrieval or planning. \textbf{PVA} adapts the PaperVizAgent workflow of PaperBanana \citep{zhu2026paperbanana} (Retriever, Planner, Stylist, Visualizer, Critic). \textbf{Static} runs \sivia{} with the initial skill $K_0$. \textbf{Memory-only} accumulates earlier findings as planning context without rewriting $K$ or $U$. \textbf{Fixed-updater} learns the skill with a frozen $U_0$. In each iteration, $U_0$ proposes two candidates, which are evaluated with the same selection and admission batches and raters as Adaptive-updater. \textbf{GEPA} \citep{agrawal2025gepa} applies reflective mutation and candidate evaluation to the skill files, without Pareto-frontier maintenance or merging, which need per-instance automatic scores that \sivia{}'s human batch judgments do not provide; it receives 172 text calls against 182 for Adaptive-updater (Table~\ref{tab:learning-system}). \textbf{Adaptive-updater} learns both the skill and the updater. Three mechanism conditions isolate the second level: \textbf{Same-parent $U_t$} and \textbf{$U_0$} apply the learned updater $U_t{=}U_2$ (after two iterations) and the initial $U_0$ to the same parent skill $K_1$ and first-iteration findings, and \textbf{Outcomes-only} replaces the edit history with outcome summaries.

\paragraph{Measures.}
Each required relation receives exactly one of five labels: correct, omitted, ambiguous, incorrect, or severe error, the last marking incorrect relations judged to misstate the mechanism; per condition the labels partition all 960 judgments (Table~\ref{tab:learning-system}). RRA (Equation~\ref{eq:rra}) counts correct labels. Two raters label every relation independently, blind to condition and in random order, and a third adjudicates disagreements (pre-adjudication agreement 85.24\%, $\kappa = 0.64$; Appendix~\ref{app:learning-evaluation}); we analyze the adjudicated labels. Six raters from the authors' research group, also blind to condition, score readability and visual quality on 1--5 scales for every family. The preference study uses 48 matched tasks from six users, eight per user, with three conditions: no profile, a fixed profile that keeps the initial rules, and a learned profile updated as in Section~\ref{sec:personal}; recording liking (1--5), binary adoption, and twenty relation labels per task. Appendix~\ref{app:learning-evaluation} reports an earlier six-rater assessment on 20 families, in which curated \sivia{} figures average 94.93\% against 87.50\% for Direct and 89.58\% for PVA (Table~\ref{tab:human-history}); the \sivia{} figures were curated from the library rather than generated by the final skill.

\paragraph{Statistics and budgets.}
Because outputs and ratings are nested within families, contrasts pair conditions by family (or task) and resample families rather than outputs \citep{dror-etal-2018-hitchhikers}. We report mean differences with 95\% intervals from 20,000 percentile-bootstrap resamples \citep{efron1979bootstrap}, together with win/tie/loss counts. Table~\ref{tab:learning-system} lists endpoint call budgets: 96 image calls per condition (288 for PVA) and 96 to 384 text calls.

\input{sections/learning-main-table}
\input{sections/learning-preference-table}

\section{Results}
\label{sec:experiments}
\input{sections/sensitivity-table}
\paragraph{Learning to draw.}
Adaptive-updater achieves 93.54\% RRA, compared with 78.23\% for Static, 84.17\% for Fixed-updater, and 85.21\% for GEPA (Table~\ref{tab:learning-system}). The family-paired differences are +15.31 $[11.77,18.96]$, +9.38 $[6.77,11.98]$, and +8.33 $[5.52,11.15]$ points, respectively. Severe errors decrease from 16 for Static to 2 for Adaptive-updater, which outperforms Static on 43 of 48 families (Figure~\ref{fig:rsi-effects}a). Direct and PVA achieve 71.46\% and 74.90\%; Adaptive-updater exceeds them by 22.08 and 18.65 points, winning on 44 of 48 families against each. Adaptive-updater also obtains readability and visual-quality scores of 4.70 and 4.78, the highest among all conditions.

\paragraph{Learning to learn.}
With a shared parent skill and identical findings, the learned updater produces a skill with 9.90 points higher RRA than $U_0$ $[6.46,13.44]$, with 34 wins and 5 losses across families (Figure~\ref{fig:rsi-effects}b); visual quality is 4.58 versus 4.10. Replacing the full edit history with outcome summaries reduces RRA by 4.17 points $[2.08,6.35]$, so the edits themselves carry information. Validation RRA increases from 73.75\% at generation 0 to 83.75\% at generation 1 and 91.67\% at generation 2 (Figure~\ref{fig:rsi-effects}c). For all five contrasts, the Bonferroni-corrected 99\% intervals exclude zero, and all leave-one-family-out estimates are positive (Table~\ref{tab:learning-sensitivity}).

\paragraph{Which errors learning removes.}
Static has 12 omitted relations among 960 labels (Table~\ref{tab:learning-system}). From Static to Adaptive-updater, incorrect relations decrease from 125 to 38 and ambiguous relations from 56 to 18. In the shared-parent comparison, skills proposed by the learned updater have 47 incorrect relations, compared with 104 for those proposed by $U_0$.

\paragraph{Personal aesthetics.}
Across 48 matched tasks, learned profiles raise mean liking from 4.02 to 4.73, +0.71 $[0.50,0.90]$ over fixed profiles (Table~\ref{tab:learning-preference}). Adoption increases from 34/48 to 45/48 tasks, and RRA from 86.35\% to 94.27\%, a gain of 7.92 points $[5.73,10.10]$. Figure~\ref{fig:profile} shows \sivia{} adapting to one individual user who prefers hand-drawn, simpler overviews.

\paragraph{Discussion.}
The gains come mainly from how relations are specified, not how many are included: planning already covers most required relations, and the remaining errors are mostly incorrect or ambiguous bindings. Under learned profiles, RRA also rises; because the contract fixes the declared structure but not the rendered image, density and whitespace affect which relations stay legible.

%% file: sections/profile-figure.tex
\begin{figure*}[t]
\centering
\begin{minipage}[t]{0.495\textwidth}
\centering
\includegraphics[width=\linewidth,trim=0 0 0 59,clip]{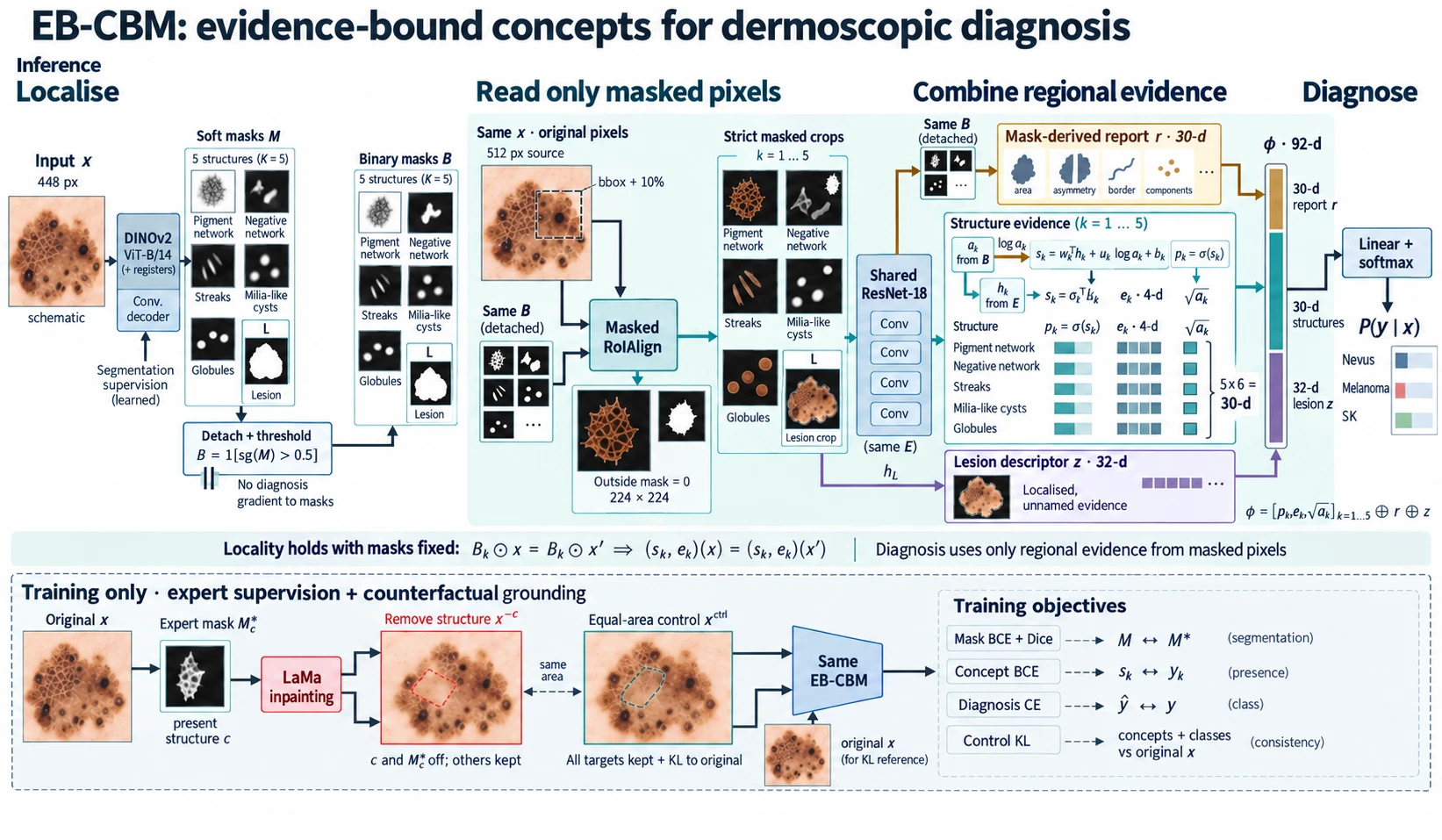}\\[1pt]
{\small (a) Fixed profile}
\end{minipage}\hfill
\begin{minipage}[t]{0.495\textwidth}
\centering
\includegraphics[width=\linewidth,trim=0 0 0 68,clip]{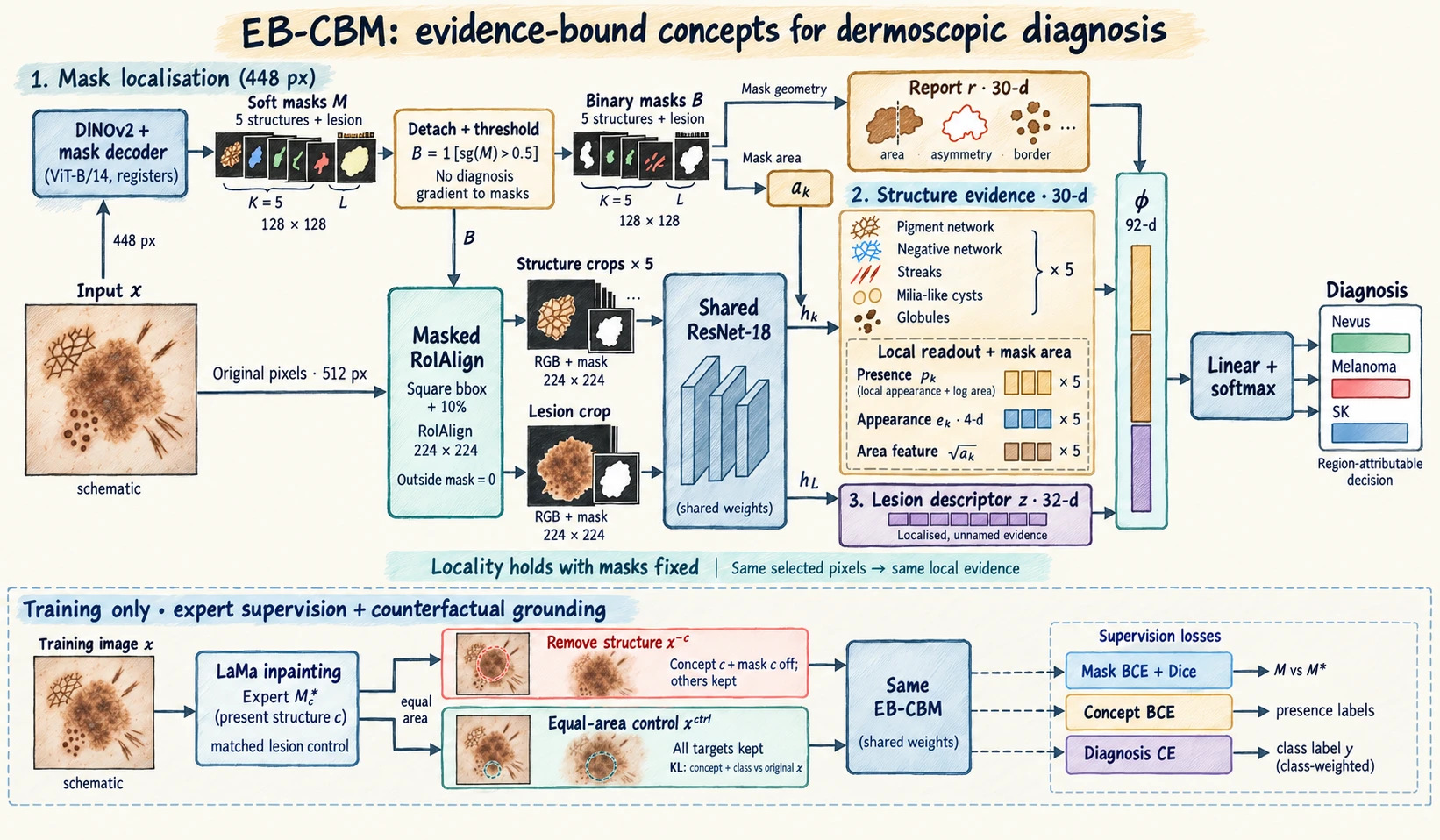}\\[1pt]
{\small (b) Learned profile}
\end{minipage}
\caption{\sivia{} overviews of a randomly selected paper. (a) Fixed profile: initial rules, dense with formulas. (b) Learned profile of one individual user who prefers hand-drawn, simpler overviews: the same modules and relations as a sketch. Profiles evolve per user; the shared skill is unchanged.}
\label{fig:profile}
\end{figure*}

%% file: sections/learning-main-table.tex
\begin{table*}[t]
\centering
\footnotesize
\setlength{\tabcolsep}{2.6pt}
\begin{tabular*}{\textwidth}{@{\extracolsep{\fill}}lccccccccccccr@{}}
\toprule
\cellcolor{SiviaHeaderBlue}Condition & \cellcolor{SiviaHeaderLilac}RRA (\%) & \cellcolor{SiviaHeaderLilac}Omit. & \cellcolor{SiviaHeaderLilac}Amb. & \cellcolor{SiviaHeaderLilac}Inc. & \cellcolor{SiviaHeaderLilac}Sev. & \cellcolor{SiviaHeaderRose}Read. & \cellcolor{SiviaHeaderRose}Visual & \cellcolor{SiviaHeaderBlue}Text & \cellcolor{SiviaHeaderBlue}Image & \cellcolor{SiviaHeaderLilac}$\Delta$RRA & \cellcolor{SiviaHeaderLilac}95\% CI & \cellcolor{SiviaHeaderRose}W/T/L\\
\midrule
\multicolumn{13}{@{}l}{\textit{External workflows}}\\
Direct & 71.46 & 14 & 75 & 166 & 19 & 3.55 & 3.60 & 96 & 96 & +22.08 & [18.65, 25.62] & 44/4/0\\
PaperVizAgent (PVA) & 74.90 & 26 & 72 & 129 & 14 & 3.80 & 3.90 & 384 & 288 & +18.65 & [15.00, 22.40] & 44/2/2\\
\addlinespace[3pt]
\multicolumn{13}{@{}l}{\textit{\sivia{} skill conditions}}\\
Static ($K_0$) & 78.23 & 12 & 56 & 125 & 16 & 3.95 & 4.00 & 144 & 96 & +15.31 & [11.77, 18.96] & 43/1/4\\
Memory-only & 79.58 & 16 & 62 & 105 & 13 & 4.05 & 4.10 & 144 & 96 & +13.96 & [10.73, 17.19] & 44/2/2\\
Fixed-updater & 84.17 & 12 & 41 & 91 & 8 & 4.15 & 4.20 & 172 & 96 & +9.38 & [6.77, 11.98] & 39/4/5\\
GEPA & 85.21 & 11 & 39 & 83 & 9 & 4.08 & 4.10 & 172 & 96 & +8.33 & [5.52, 11.15] & 31/12/5\\
\textbf{Adaptive-updater} & \textbf{93.54} & \textbf{4} & \textbf{18} & \textbf{38} & \textbf{2} & \textbf{4.70} & \textbf{4.78} & 182 & 96 & \multicolumn{3}{c}{\textit{baseline}}\\
\addlinespace[3pt]
\multicolumn{13}{@{}l}{\textit{RSI mechanism conditions}}\\
Same-parent $U_0$ & 81.56 & 16 & 47 & 104 & 10 & 3.98 & 4.10 & 182 & 96 & +9.90$^{\dagger}$ & [6.46, 13.44] & 34/9/5\\
Same-parent $U_t$ & 91.46 & 5 & 26 & 47 & 4 & 4.47 & 4.58 & 182 & 96 & \multicolumn{3}{c}{\textit{baseline}$^{\dagger}$}\\
Outcomes-only & 89.38 & 8 & 30 & 58 & 6 & 4.36 & 4.46 & 182 & 96 & +4.17 & [2.08, 6.35] & 26/15/7\\
\bottomrule
\end{tabular*}
\caption{Main results on the 48 test families (960 relation labels per row). Omit., Amb., Inc., and Sev.\ count omitted, ambiguous, incorrect, and severe-error labels; Read.\ and Visual average six raters' 1--5 scores and are reported descriptively, without intervals; Text and Image are endpoint call budgets. $\Delta$RRA is the family-paired gain of Adaptive-updater over the row ($^{\dagger}$of Same-parent $U_t$ over $U_0$) with 95\% bootstrap intervals; W/T/L counts families where Adaptive-updater is higher, tied, or lower.}
\label{tab:learning-system}
\label{tab:learning-paired}
\end{table*}

%% file: sections/learning-preference-table.tex
\begin{table}[t]
\centering
\small
\setlength{\tabcolsep}{3.5pt}
\begin{tabular}{@{}lcccr@{}}
\toprule
\cellcolor{SiviaHeaderBlue}Profile & \cellcolor{SiviaHeaderRose}Liking & \cellcolor{SiviaHeaderRose}Adopted & \cellcolor{SiviaHeaderLilac}RRA (\%) & \cellcolor{SiviaHeaderLilac}Sev.\\
\midrule
None (shared) & 3.54 & 23/48 & 80.10 & 7\\
Fixed & 4.02 & 34/48 & 86.35 & 4\\
Learned & 4.73 & 45/48 & 94.27 & 1\\
\bottomrule
\end{tabular}
\caption{Personal adaptation on 48 matched tasks: liking (1--5), binary adoption, RRA over twenty relations per task, and severe-error labels (Sev.).}
\label{tab:learning-preference}
\end{table}

%% file: sections/sensitivity-table.tex
\begin{table}[t]
\centering
\small
\setlength{\tabcolsep}{3pt}
\begin{tabular}{@{}lcr@{}}
\toprule
\cellcolor{SiviaHeaderBlue}Contrast & \cellcolor{SiviaHeaderLilac}99\% interval & \cellcolor{SiviaHeaderRose}\shortstack{Leave-one-\\family-out}\\
\midrule
Adaptive $-$ Static & [10.63, 20.10] & [14.68, 15.74]\\
Adaptive $-$ Fixed & [6.04, 12.81] & [8.83, 9.79]\\
Adaptive $-$ GEPA & [4.69, 12.19] & [7.77, 8.72]\\
Same-parent $U_t-U_0$ & [5.52, 14.58] & [9.15, 10.43]\\
Adaptive $-$ outcomes-only & [1.46, 7.08] & [3.72, 4.57]\\
\bottomrule
\end{tabular}
\caption{Robustness of the five learning contrasts (RRA points): 99\% bootstrap intervals and the range of leave-one-family-out means.}
\label{tab:learning-sensitivity}
\end{table}

%% file: sections/conclusion.tex
\section{Conclusion}
An overview figure must be both accurate and beautiful. \sivia{} combines source-derived relation planning with human-guided RSI, revising both its drawing skill and the updater that proposes skill edits. On 48 test papers, learning both levels raises RRA to 93.54\% and readability and visual quality to 4.70 and 4.78 of 5, the highest of all conditions, and from a shared parent skill the learned updater's proposal gains 9.90 points. Personal profiles adapt presentation to their user's taste, raising liking from 4.02 to 4.73 and adoption from 34 to 45 of 48 tasks without changing the scientific contract; Figure~\ref{fig:profile} shows this adaptation for one individual user.

\section*{Limitations}
All learning results come from a single learning history, so run-to-run variation is not estimated; the same-parent comparison uses one parent skill, one text model, and one image generator. Relation labels are produced by three raters, and visual ratings come from six raters in the authors' research group, who also selected and admitted updates; the preference study involves six users. RRA scores the required relations of each checklist. Endpoint budgets cover the final evaluation, and the GEPA baseline omits Pareto-frontier maintenance and merging. The diagnostic study is likewise assessed by three raters, and all studies concern overview figures of machine-learning papers.

\section*{Ethical Considerations}
A convincing overview figure can still depict an unsupported mechanism. \sivia{} therefore anchors every relation to passages of the source paper and assesses scientific accuracy separately from visual appeal; authors should still check each figure against their paper. AI tools assisted with implementation, analysis, literature search, and drafting.

%% file: sections/appendix.tex
\FloatBarrier

\section{Illustration Gallery}
\label{app:gallery-section}
Figures~\ref{app:gallery}--\ref{app:gallery-last} show six curated \sivia{} overview figures for published methods in multi-view geometry, video reconstruction, voxel interpretation, visual question answering, and retrieval-augmented generation.
\input{sections/gallery}

\section{Implementation Details}
\label{app:implementation}
Skill edits are bounded replacements in the skill's text files. Each completed iteration stores both candidates' before/after edits, their producing updaters, the source families, the scientific assessments and preferences, and both human decisions.

The drawing workflow resolves the source, the template, and any applicable preferences and compiles the final image-generation request; a separate research runner manages learning, call budgets, and human review. Each candidate derives its scientific contract from the target source with its own skill and never sees the evaluation checklist. Personal-profile comparisons instead fix both the shared skill and the contract and vary only presentation rules. Memory-only adds earlier confirmed scientific and readability findings, without personal notes or validation feedback, to the planning input of $K_0$; the final evaluation uses the memory at the end of learning.

The released artifact at \url{\repoURL} packages the same-parent comparison used in Table~\ref{tab:learning-system}: the per-relation labels, per-output correct and severe-error counts, six-rater readability and visual-quality scores, endpoint budgets, and the parent-skill and updater configurations. Other conditions are not included.

\section{Diagnostic Study Details}
\label{sec:results}
This appendix describes the corpus, generation inputs, and annotation of the diagnostic study in Section~\ref{sec:diagnostic}.
\subsection{Protocol and Sensitivity}
\label{app:diagnostic}
\paragraph{Corpus construction.}
We select 445 papers from over 5,000 candidates, focusing on leading-venue research from the preceding two years while retaining earlier studies such as Tree of Thoughts \citep{yao2023tot}. Each selected paper is one evaluation family and contributes multiple conditions assessed against shared source-grounded requirements: 1,964 generated images, 4,033 distinct required relations, and 17,973 human image--requirement judgments in total. 

The development cohorts span multi-agent systems, verification and reflection, tool control, retrieval and memory, reasoning and search, and learning and adaptation. An assistant assigned paper types from titles and abstracts: methodological (271), non-language-centric (97), benchmark or dataset (42), and analysis (35).

\paragraph{Generation and retrieval.}
Cohort A retrieves over 948 papers using TF--IDF; B and C use 781 papers and \texttt{all-mpnet-base-v2} embeddings. Title-and-abstract queries retrieve architecture, framework, or pipeline diagrams, excluding known target identities and aliases. B's random-reference condition shares the eligible pool and wording. These conditions add one image to shared generation inputs.

Cohort~A schedules three conditions with two repetitions, Cohort~B four with two, and Cohort~C three with one. C's first 40 families use the original layout; the other 255 share a compact-layout instruction and revised whitespace rubric. Conditions within a family share instructions. Bootstrap resampling preserves stratum sizes; layout, paper composition, and generation time vary together across strata.

\paragraph{Checklists and review.}
Pre-generation requirements specify source anchors and minimum visible evidence. Requirements are drafted by a separate language-model session. Checklists have 5--8, 7--14, and 5--12 items, respectively. Compound items require all components to be visible. The same frozen requirements apply to all images of a family. Relation-plan extraction is separate from checklist preparation.

Review hides conditions and prompts, shuffles paper blocks and images, and permits source-support objections. Three raters share the assessment workload.

\paragraph{Aggregation and sensitivity.}
Contrasts require all prescribed outputs in both conditions. Scores average within family and condition, then weight families equally. Intervals use 10,000 family-level bootstrap resamples, stratified by layout in Cohort~C.

Source objections affect 30 requirements in 27 families. Excluding disputed items across all family images gives Plan$-$Brief gains of +4.26, +14.38, and +19.85 points in Cohorts~A, B, and~C; averaging available repetitions gives +4.38, +14.54, and +19.85. The main text and Figure~\ref{fig:scatter} report the all-outputs convention: +4.38, +14.52, and +19.85.

\subsection{Two Failure Cases}
\label{app:diagnostic-cases}
Mobile-Agent-RAG requires perception after a mobile action \citep{zhou2026mobileagentrag}. Its source-derived plan and generation prompt preserve that requirement. One generated image routes the updated screenshot onward without the post-action perception operation; a second image from the same prompt makes both pre-action and post-action perception explicit. Their frozen-checklist scores are 7/8 and 8/8. The distinction locates the first error in realization, after the requirement has entered the prompt.

AI-Researcher illustrates both role binding and plan coverage \citep{tang2025airesearcher}. The source assigns atomic-concept decomposition to resource analysis; an inspected image retains a later decomposition operation but omits this earlier actor--operation binding. Separately, checking code against an implementation plan is not explicitly preserved in the extracted plan, which describes using the plan to construct code. These cases direct feedback toward different changes: strengthen visible endpoint and phase instructions when the specification is already present, or improve source-to-plan extraction when it is absent.

\section{Learning Evaluation Details}
\label{app:learning-evaluation}
This appendix supplements Sections~\ref{sec:setup} and~\ref{sec:experiments}.

\input{sections/learning-history-table}
\input{sections/learning-evaluation-details}

%% file: sections/gallery.tex
\begin{figure*}[p]
\centering
\includegraphics[width=\textwidth,height=0.38\textheight,keepaspectratio]{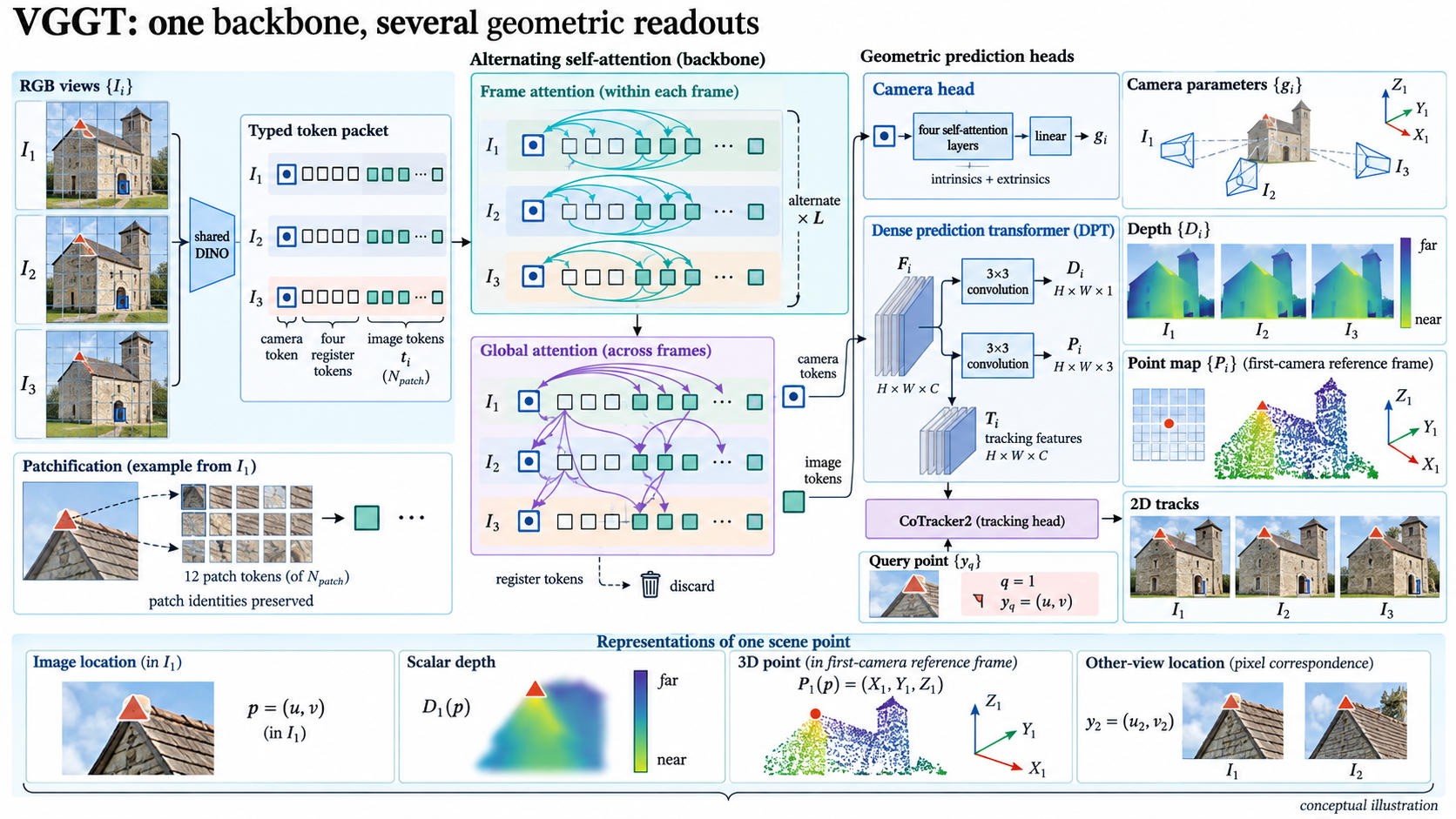}
\caption{VGGT \citep{wang2025vggt}: shared attention and task-specific heads for multi-view geometry.}
\label{app:gallery}
\medskip
\centering
\includegraphics[width=\textwidth,height=0.38\textheight,keepaspectratio]{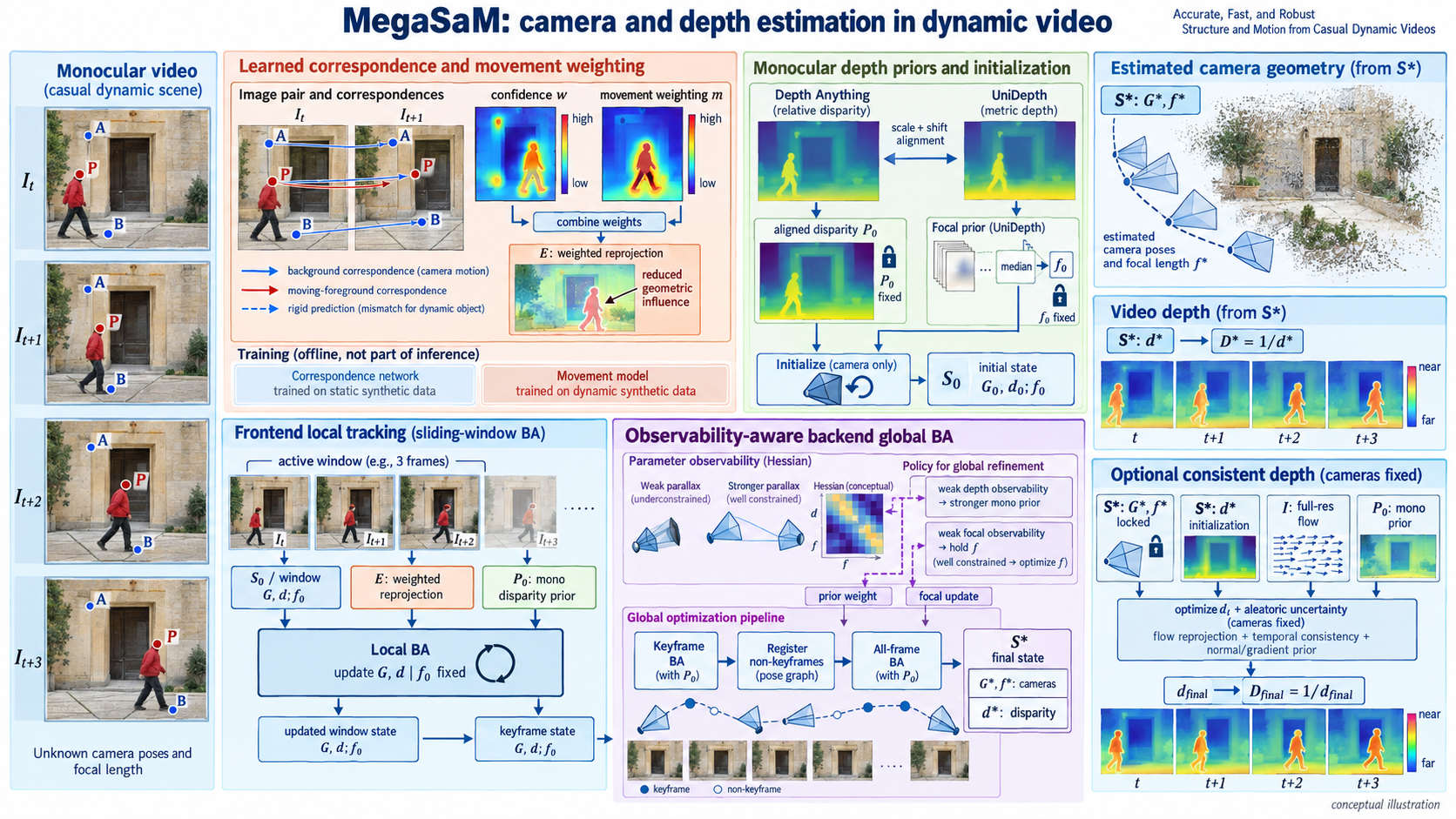}
\caption{MegaSaM \citep{li2025megasam}: camera and depth estimation in dynamic video.}
\end{figure*}

\begin{figure*}[p]
\centering
\includegraphics[width=\textwidth,height=0.38\textheight,keepaspectratio]{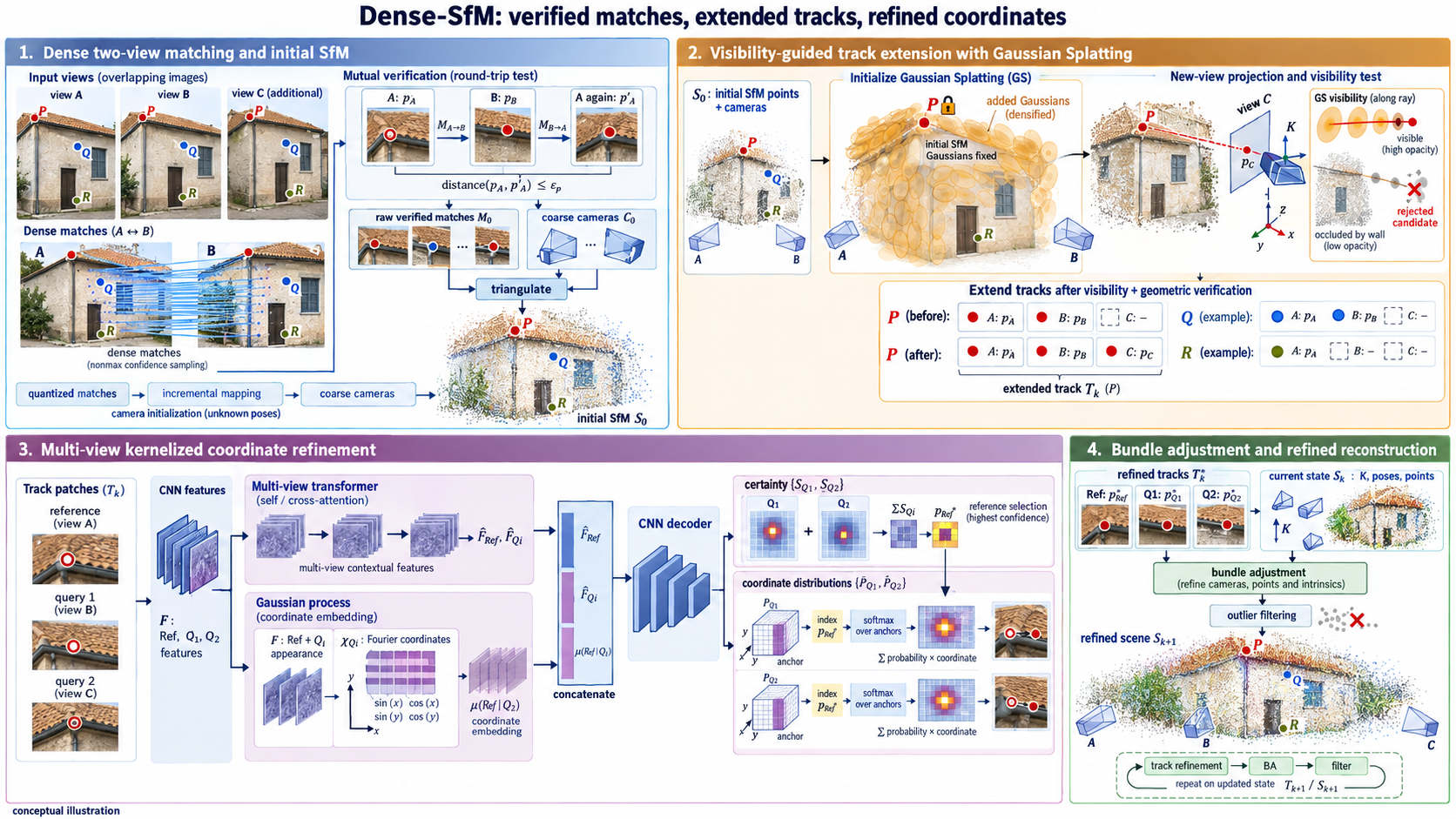}
\caption{Dense-SfM \citep{lee2025densesfm}: verified matches, visibility-guided track extension, and coordinate refinement.}
\medskip
\centering
\includegraphics[width=\textwidth,height=0.38\textheight,keepaspectratio]{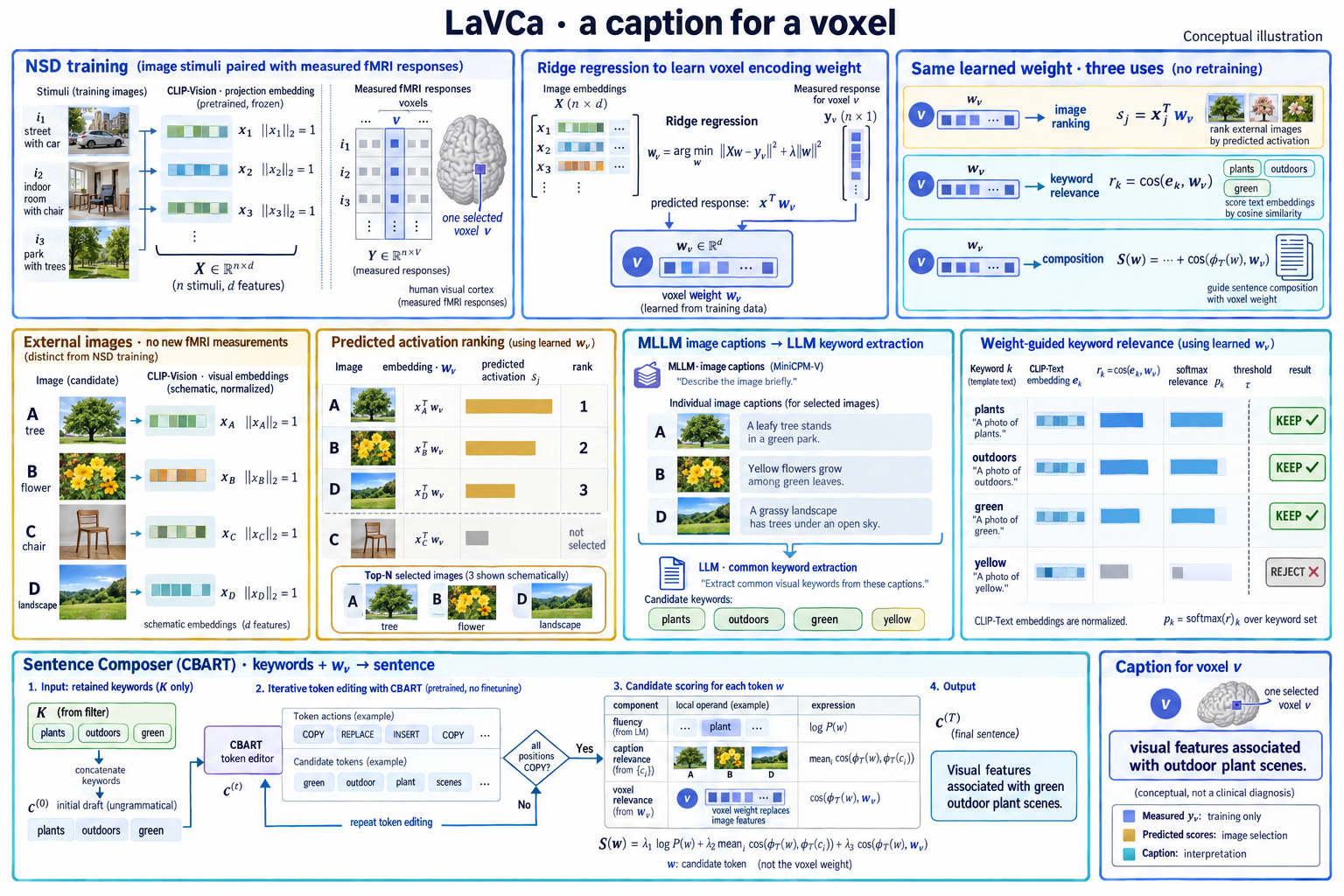}
\caption{LaVCa \citep{matsuyama2026lavca}: voxel interpretation through image ranking, keyword selection, and caption composition.}
\end{figure*}

\begin{figure*}[p]
\centering
\includegraphics[width=\textwidth,height=0.38\textheight,keepaspectratio]{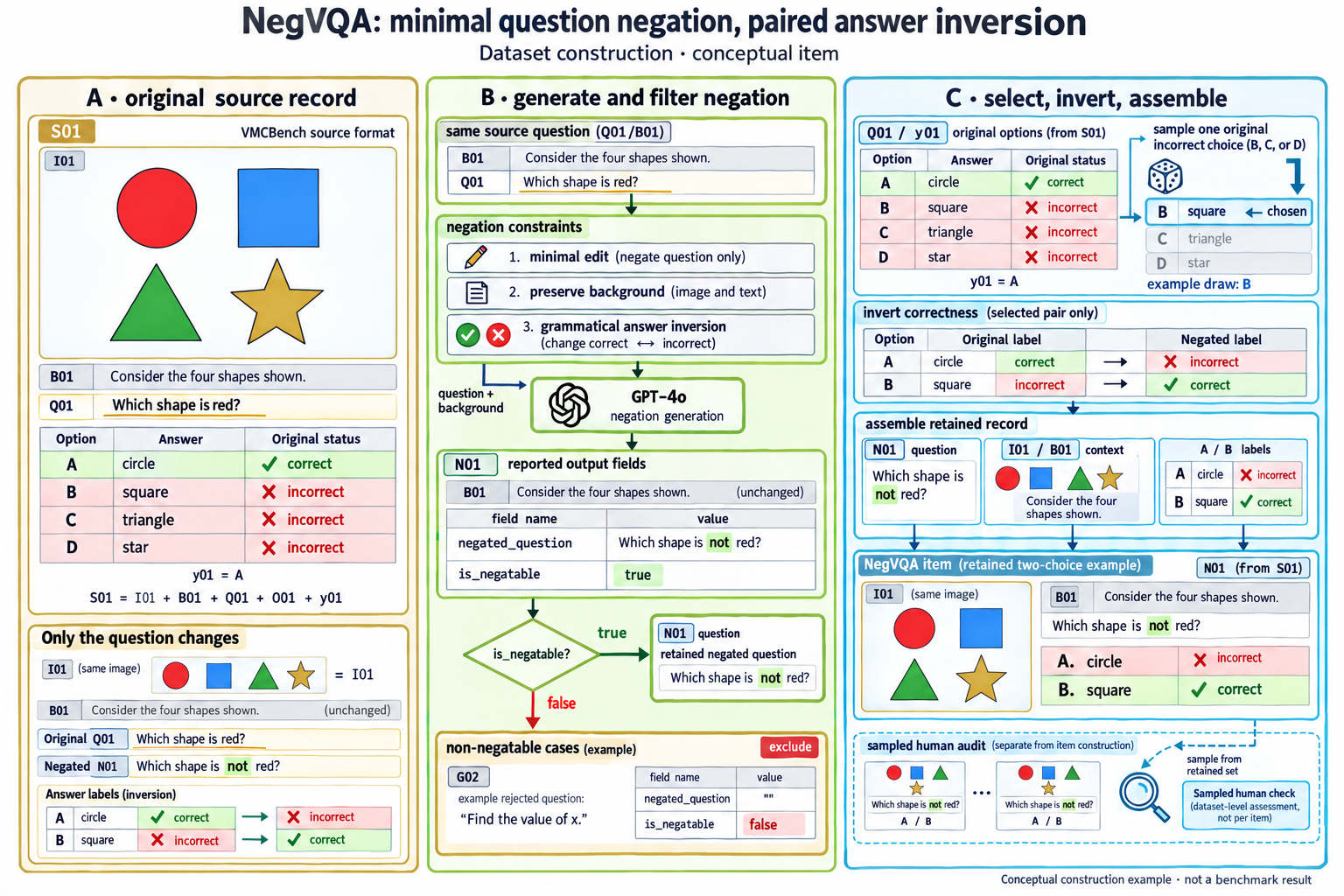}
\caption{NegVQA \citep{zhang2025negvqa}: minimal question negation and paired answer inversion for visual question answering.}
\medskip
\centering
\includegraphics[width=\textwidth,height=0.38\textheight,keepaspectratio]{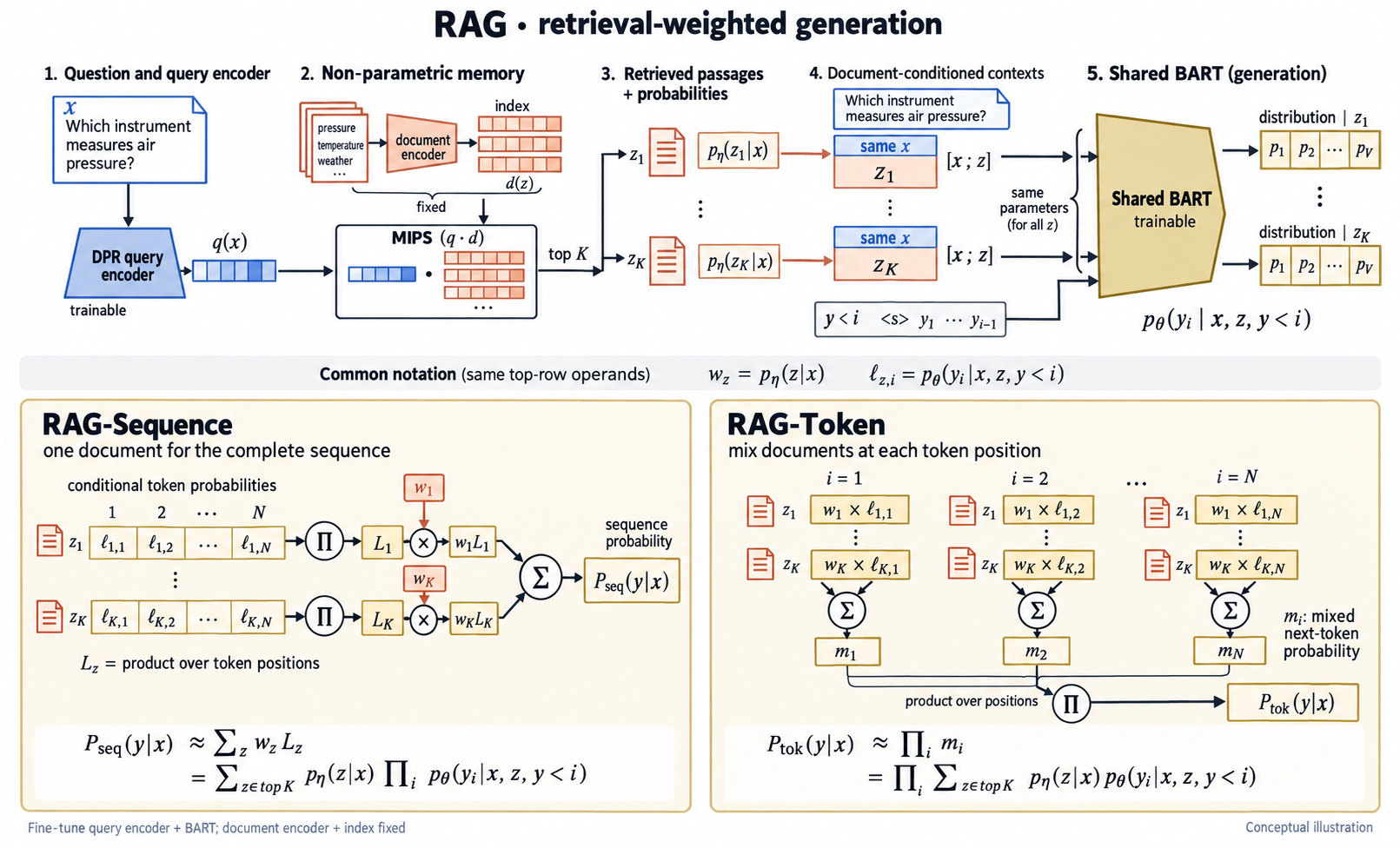}
\caption{RAG \citep{lewis2020rag}: retrieval-weighted generation with sequence-level and token-level document marginalization.}
\label{app:gallery-last}
\end{figure*}

%% file: sections/learning-history-table.tex
\begin{table*}[t]
\centering
\small
\setlength{\tabcolsep}{2pt}
\begin{tabular*}{\textwidth}{@{\extracolsep{\fill}}lccccccccr@{}}
\toprule
\cellcolor{SiviaHeaderBlue}Method & \cellcolor{SiviaHeaderLilac}Correct/total & \cellcolor{SiviaHeaderLilac}C/A (\%) & \cellcolor{SiviaHeaderRose}Reading & \cellcolor{SiviaHeaderRose}Hier. & \cellcolor{SiviaHeaderRose}Readab. & \cellcolor{SiviaHeaderRose}Layout & \cellcolor{SiviaHeaderRose}Aesth. & \cellcolor{SiviaHeaderRose}Liking & \cellcolor{SiviaHeaderRose}Adopted\\
\midrule
Direct first & 153/175 & 87.50 & 3.90 & 4.05 & 3.85 & 3.80 & 3.85 & 3.80 & 68/120\\
PVA final & 157/175 & 89.58 & 3.95 & 4.10 & 3.90 & 3.75 & 3.90 & 3.95 & 76/120\\
Sivia library & 166/175 & 94.93 & 4.50 & 4.60 & 4.40 & 4.55 & 4.65 & 4.70 & 111/120\\
\bottomrule
\end{tabular*}
\caption{Historical human assessment of curated \sivia{} figures and two baselines on 20 families. C/A is averaged over families. Each relation receives six votes and each family six visual and preference ratings (1--5); Hier., Readab., and Aesth.\ denote hierarchy, readability, and overall aesthetics, and adoption is binary.}
\label{tab:human-history}
\label{tab:human-history-preference}
\end{table*}

%% file: sections/learning-evaluation-details.tex
\paragraph{Families and conditions.}
The test set has 48 families, each with two outputs per condition and ten required relations per output: 6,720 relation labels for the seven main conditions and 2,880 for the three mechanism conditions. The 24 validation families, which are disjoint from the test families, have two outputs at each of generations 0--2. The preference study has 48 tasks, three conditions, and twenty relations per task.

\paragraph{Labels and aggregation.}
Two raters label every relation independently, blind to condition and in random order, and a third adjudicates disagreements; on a pre-designated panel of the 24 odd-numbered test families across the seven main conditions (3,360 relations), their pre-adjudication labels agree on 85.24\% of relations (Cohen's $\kappa = 0.64$). Each relation receives one of five labels: correct, omitted, ambiguous, incorrect, or severe error, the last marking an incorrect relation that misstates the mechanism. Table~\ref{tab:learning-system} lists the counts. RRA divides an output's correct relations by its ten requirements, averages the two outputs of a family, and then averages families. Six raters, blind to condition, score readability and visual quality on 1--5 scales for every family and condition (48 $\times$ 6 ratings per condition); scores are averaged within and then across families.

\paragraph{Paired uncertainty.}
Contrasts pair the same families across conditions. Because outputs and ratings are nested within families, we resample families \citep{dror-etal-2018-hitchhikers}: we draw 20,000 bootstrap samples \citep{efron1979bootstrap} of the 48 family differences and report the 2.5th and 97.5th percentiles.

\paragraph{Robustness.}
\label{app:learning-sensitivity}
Table~\ref{tab:learning-sensitivity} reports 99\% intervals from the same bootstrap draws, which correspond to a Bonferroni correction over the five learning contrasts, and the range of each contrast when one family is omitted at a time.

\paragraph{Personal aesthetics.}
Learned-minus-fixed differences pair the 48 tasks and use 20,000 task-level bootstrap resamples. The tasks come from six users, eight per user.

\paragraph{Historical assessment.}
Table~\ref{tab:human-history} covers 20 families for each of Direct, PVA, and curated \sivia{} figures, with 175 relations per method. Six raters vote on every relation, and we use the majority label, which is supported by at least four of six votes in every case. C/A averages each family's fraction correct and therefore differs slightly from pooling all 175 relations. Liking averages the 120 ratings per method, and adoption counts positive responses.

\paragraph{Budgets.}
Each condition has 96 image calls (288 for PVA) and 12.8 review hours. Text calls are 96 for Direct, 384 for PVA, 144 for Static and Memory-only, 172 for Fixed-updater and GEPA, and 182 for Adaptive-updater and the mechanism conditions.